\documentclass[fleqn,usenatbib,floatfix]{mnras}
\usepackage{longtable}
\usepackage{amssymb}
\usepackage{amsmath}
\usepackage{multirow}
\usepackage{multicol}
\usepackage{longtable}
\usepackage{hyperref}
\usepackage{graphicx}
\usepackage{dcolumn}
\usepackage{bm}
\usepackage[utf8x]{inputenc}
\usepackage{url}
\usepackage{comment}
\usepackage{verbatim}
\usepackage{newtxtext,newtxmath}
\usepackage[T1]{fontenc}

\DeclareRobustCommand{\VAN}[3]{#2}
\let\VANthebibliography\thebibliography
\def\thebibliography{\DeclareRobustCommand{\VAN}[3]{##3}\VANthebibliography}

\usepackage{graphicx}	
\usepackage{amsmath}	
\usepackage{amssymb}	
\usepackage{gensymb}

\usepackage{array}
\usepackage{multirow}
\usepackage{longtable}
\usepackage{tabu}
\usepackage{supertabular,booktabs}

\title{Estimating dynamical parameters of two interacting galaxies using Deep Learning}

\author[]{
Adarsh Mahor$^{1,a}$\thanks{\href{mailto: u19me202@med.svnit.ac.in}{ u19me202@med.svnit.ac.in}}$\dagger$,
Janvita Reddy$^{1,a}$\thanks{Both authors have an equal amount of contribution},
Amitesh Singh$^{2,a}$,
Shashwat Singh$^{3,a}$
\\
$^{1}$Sardar Vallabhbhai National Institute of Technology, Surat, Gujarat, India-395007\\
$^{2}$University of Mississippi, University, MS 38677, USA \\
$^{3}$Sorbonne University, UPMC, Paris, France  \\
$^{a}$Bose.X TRIAC Collaboration\thanks{\url{https://www.bosex.org/}} \\
}

\date{Accepted XXX. Received YYY; in original form ZZZ}

\pubyear{2021}

\begin{document}
\label{firstpage}
\pagerange{\pageref{firstpage}--\pageref{lastpage}}
\maketitle

\begin{abstract}

The science behind galaxy interaction and mergers has a fundamental role and gives us an insight into galaxy formation and its evolution. Fluctuating angular momentum is responsible for extraordinary events like polar rings, tidal tails, and ripples. To study different phenomena related to galaxy interactions, various parameters like the mass ratio of the interacting galaxy, orbital parameters, mass distribution, morphologies are required. Convolutional Neural Networks (CNN) are widely used to classify image data. Thus, we used CNN as our approach to the problem. In this work, we will be using data from state-of-the-art magneto-hydrodynamic simulations of galaxy mergers from the GalMer database at different dynamical parameters using image snapshots of merging pairs of galaxies and feeding them to our Deep Learning model (ResNet). The dynamical parameters we are aiming for; would be spin, relative inclination ($i$), viewing angle ($\theta$), and azimuthal angle ($\phi$). We aim to download bulk data using the web scraping method. The first approach is to create different combinations of these parameters to form 60 classes. Feeding the data into the model, we achieved 93.63\% accuracy. As we received good results in minute classification, we moved to our second approach, regression. Here the model can predict the continuous and exact values of the dynamical parameters. We have achieved a 99.86\% R-squared value and the mean squared error of 0.0833 on testing data. In the end, we used data from Sloan Digital Sky Survey to test our trained model on some real images. 

\end{abstract}




\section{Introduction}
\label{Introduction}

Edwin Hubble was the first to give a galaxy classification scheme in 1926, known as the Hubble tuning-fork diagram. \citet{abraham2000explorations} have developed a quantitative two-parameter description of galactic structure. In the modern cosmological model, the universe formation has an interacting galaxy as its main element. This cosmological model also includes the formation of massive galaxies due to the interaction and merger of multiple dwarf galaxies founded by the abundance of early-type galaxies are given on the left side of the Hubble tuning-fork diagram at higher red shifts and have high rates of galaxy interaction. Galaxy
interactions are even responsible for dynamic evolution and regulate some astronomical events like morphology variations, formation of bulges, nuclear activities, creations of halo, starbursts, and many more. Very few of them evolve in the form of steady evolution \citet{barnes1992dynamics}.

\citet{holmberg1937study} did one of the path-finder works in the field of interacting galaxies. \citet{vorontsov1959atlas} published a catalog of interacting galaxies in the astronomical council of the USSR academy of sciences. The record and atlas contain 852 interacting galaxies. The first part was published in 1959. This catalog contains 355 interacting galaxies numbered $VV1$ through $VV355$. The second part was published in the 1970s and included interacting galaxies numbered $VV356$ through $VV852$.\citet{vorontsov2003vizier} added additional 1162 objects ranging from $VV853$ to $VV2014$ from the Morphological Catalogue of Galaxies (MCG) or Morfologiceskij Katalog Galaktik. It is a Russian catalog of 30,642 galaxies anthologized by Boris Vorontsov-Velyaminov and V. P. Arkhipova. It is based on the scrutiny of the Palomar Observatory Sky Survey (POSS). \citet{hibbard2001hi} published a catalog of peculiar galaxies interacting at different wavelengths, and \citet{arp1966atlas} studied queerness and deformations in 338 interacting pairs

Interaction of the galaxies is a complicated dynamical issue in astronomy, and generally, it is not logically tractable. Using an optical analog computer to perform N-body integrations, \citet{holmberg1941clustering} investigated whether the tidal disturbances cause energy in an interacting pair of galaxies. The author unwillingly rejected the idea that repeated tidal encounters would cause galaxies to merge. Numerous interactions in close neighborhoods can result in galaxies interference was first proposed by \citet{zwicky1956multiple}. The same possibility is explained by \citet{alladin1965dynamics} using hyperbolic encounters of spheroidal galaxies. Based on close encounters in a merging pair, prolate ellipsoidal and medium hyperbolic orbits for the motion of a secondary match are considered by \citet{yabushita1971possibility} and \citet{tashpulatov1969tidal}. In \citet{toomre1972galactic}, they show the bridges and tails seen in multiple galaxies are just tidal relics of close encounters. Here they considered parabolic encounters to form bridges and tails in galaxy mergers. \citet{toomre1972galactic} employed only some hundreds of particles to formulate dynamical models of pairs of galaxy interaction, so these simulations are considered low-resolution simulations. Afterward, researchers mainly simulated the galactic disc as a stellar disc, which is self-gravitating through the galactic disc is a multi-component system of stars, gravitationally coupled system, gas, and the dark matter halo mentioned in \citet{bodenheimer2006numerical}. 

Choice of initial conditions for interacting pairs of galaxies firmly constructs a dynamic model. Especially models are highly dependent on the observational constraint. 
For the initial conditions, someone may require many dynamical parameters associated with orbital geometry, energy, spin, and mass ratios. In addition, other sets of parameters like 
velocity scales, length, distances, and viewing directions have to be chosen to fit the model results with the observed structure and kinematics (\citet{toomre1972galactic}; \citet{barnes2009identikit}; \citet{chilingarian2010galmer}; \citet{barnes2011identikit}; \citet{privon2013dynamical}; \citet{mortazavi2015modeling}). The trial and error method is used to know the interacting galaxies' initial condition. But N-body and hydrodynamical simulations are computationally expensive, so this method is not practically applicable. However, \citet{bekki2019constraining} came up with an application on Deep Convolutional Neural Network (DCNN) on Smooth Particle Hydrodynamics (SPH) simulation data by constraining the three-dimensional orbits of galaxies under Ram Pressure Stripping (RPS) to determine the orbital geometry of satellite galaxies in galaxy clusters.

\citet{prakash2020determination} is the inspiration behind our work. His paper has shown a method to find parameters of the two interacting galaxies parameters. Their writing has offered to find the viewing angle and relative inclination by forming a classification of groups. In this paper, we build a model that can determine dynamical parameters, such as the viewing angle and relative inclination, including other parameters. Presently, machine learning techniques have been applied to various problems in almost every field. In astronomy, we use machine learning to classify or predict any astrophysical object, event, or context as the classification of stars and galaxies. \citet{weir1995automated} present an experimental study of the performance of three machine learning algorithms applied to the complex problem of galaxy classification. \citet{conselice2006fundamental} shows the paper with a new three-dimensional galaxy classification system designed to account for the diversity of galaxy properties in the nearby universe. Optical transient events are shown in \citet{cabrera2017deep} and \citet{mahabal2011discovery}. Rotation-invariant CNNs for galaxy morphology prediction \citet{dieleman2015rotation}. For classifying radio galaxy images \citet{aniyan2017classifying} used data from the Very Large Array (VLA) and achieved an accuracy of 95\%. \citet{abraham2018detection} achieved 94\% accuracy for classifying barred and un-barred galaxies. \citet{flamary2017astronomical} used CNN to reconstruct astronomical images, which provided an efficient model in terms of reconstruction and computational speed. \citet{dai2018galaxy} performed galaxy morphology classification with DCNN, and the overall classification accuracy of the network was 95\%. Recently \citet{jernelv2020convolutional} used CNN for classification and regression analysis of one-dimensional spectral data.

The paper is structured as follows. Section \ref{Galmer Database:2}, describes the GalMer database and the schematic representation of orbital geometry in detail. Section \ref{CNNs in dynamical parameter estimation:3} explains the significance and estimation of dynamical parameters using DCNN. In Section \ref{Data collection and preprocessing:4} method for data collection and preprocessing is described in brief. The CNN architecture we used is explained in Section \ref{Convolutional Neural Networks:5}, and results obtained by these Models are discussed in Section \ref{Results}. We finally summarise this work and conclude in Section \ref{Conclusion}.  

\section{GalMer Database}
\label{Galmer Database:2}

The GalMer project is related and developed in the French national HORIZON framework collaboration has the ambitious goal of providing access for the astronomical community to the results of high and moderate resolution numerical simulations of galaxy interactions in pairs. They tried to cover the parameter space of the initial conditions as much as possible, thus allowing to study star formation enhancements, structural and dynamical properties of merger remnants statistically. The GalMer database is a library of thousands of simulations of galaxy mergers at moderate spatial resolution. It is a compromise between the diversity of initial conditions and the details of underlying physics. GalMer is an N-body + Smoothed-particle hydrodynamics galaxy merger simulation with models which consist of a non-rotatable dark matter halo, which shall or shall not include a gaseous and a stellar disc, and alternatively, a central non-rotating bulge (\citet{chilingarian2010galmer}). For every pair of galaxies, they set the inclination ($i_1$) value of the one disc to 0°, and the inclination ($i_2$) value of the other disc is varied from 0°, 45°, 75°, and 90°. But for giant-dwarf interactions, the inclination $i_1$ = 33°and $i_2$ = 130° and set as default for the more generic case. 

\begin{figure}
\includegraphics[width = 0.55\textwidth]{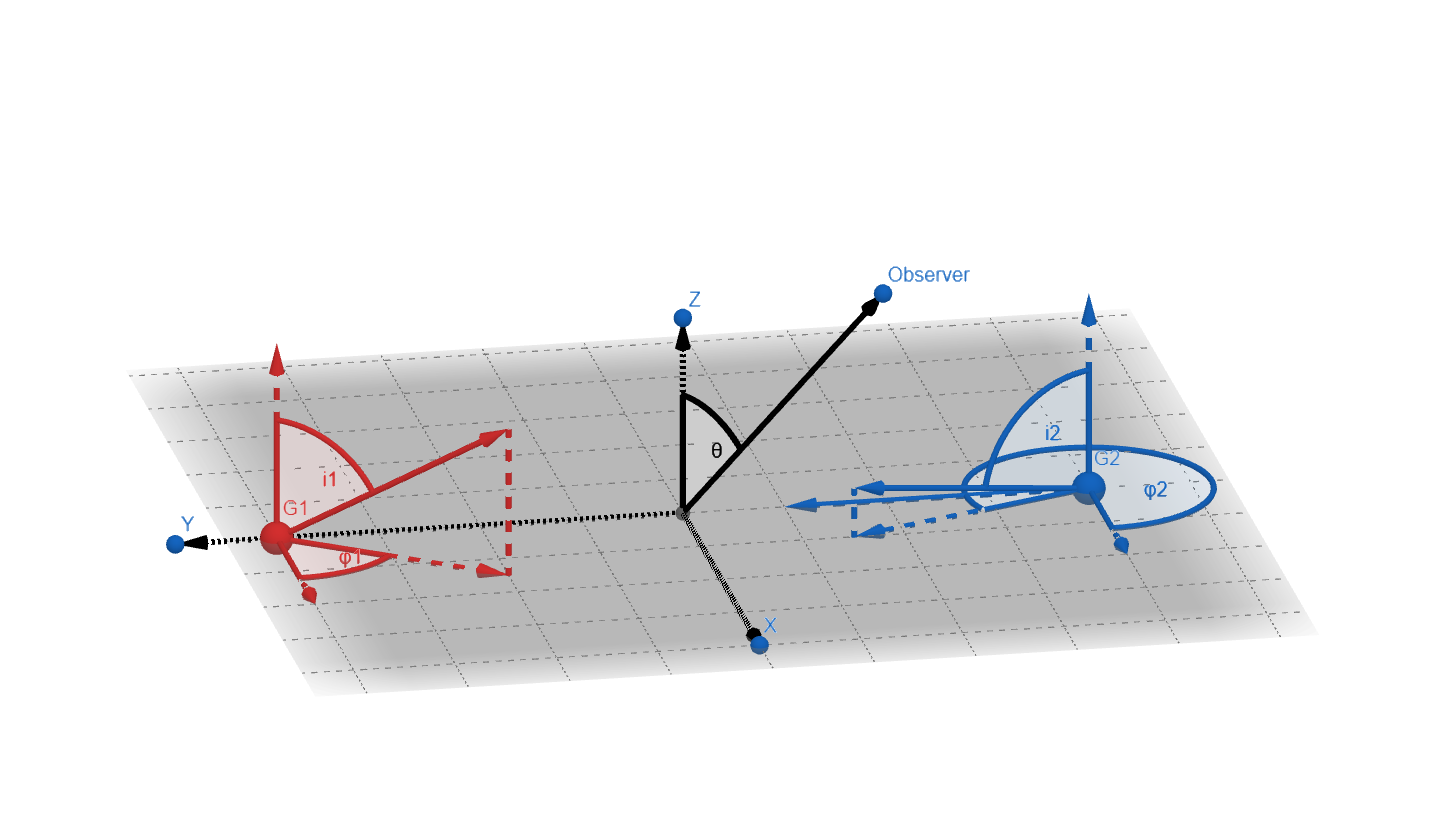}
\caption{Representation of Cartesian coordinate system present in GalMer Database. G1 and G2 represent the two interacting galaxies while $i_1$ and $i_2$ are the angle of inclination of respective galaxies. $\theta$ represent the viewing angle of the two interacting galaxies. $\phi_1$ and $\phi_2$ represent azimuthal angles of the corresponding two galaxies. Relative inclination ($i$) is given by $i_2 - i_1$.}
\label{fig:Galmer(8)}
\end{figure}

In fig \ref{fig:Galmer(8)}, we represent adopted orbital geometry for the simulation. We have set up the collision so that the orbital angular momentum is parallel to the z-axis and that the centers of the two galaxies are initially on the x-axis. The normal to the orbital plane coincides with the z-axis of the 3D cartesian coordinate system.
The angle subtended by the perpendicular to the orbital plane and the line of sight concerning the observer is known as the viewing angle ($\theta$), and it ranges from 0° to 90°. G1 and G2 denote the interacting pair of galaxies. Galaxy spins are specified in terms of the spherical coordinates ($i_1,\phi_1$) and ($i_2$, $\phi_2$) as such, the angle between the orbital motion, which is z-axis and the axis of spin are given by $i_1$ and $i_2$ for respective G1 and G2 galaxy. Azimuthal angles are between the spin axis's projection on the orbital plane, and the x-axis is denoted by $\phi$. It ranges from 0° to 180°. In GalMer Simulations, the angle of inclination of the first galaxy ($i_1$) is taken as zero, which means it lies in the orbital plane. Consequently, the difference between the inclination of galaxies, i.e., $i_2 - i_1$, is known as the relative inclination ($i$) of an interacting galaxy pair, equal to the inclination of the second galaxy ($i_2$). The probability of the spin $i_2$ of the second galaxy to be oriented between 0 and $i_2$ is proportional to 1−cos($i_2$). 

\section{CNN in dynamical parameter estimation and it's significance}
\label{CNNs in dynamical parameter estimation:3}

For estimation of dynamical parameters like spin, $i$, $\theta$, and $\phi$; we are using DCNN. ($i$) can have a continuous value in the range of 0° to 90°, but the GalMer database simulates only four discrete sets of values for the 1:1 mass ratio case. Those are 0°, 45°, 75°, and 90°, But for each ($i$), $\theta$ and $\phi$ can range from -90° to 90° and -180° to 180° respectively. Spin, the dynamical parameter, has two fixed values: Prograde or Retrograde. We have considered only images of orbit-type-1. 
The reason behind choosing this orbit was the pericentral distance, which is 8 Kpc, and motion energy is 0, proving advantageous as giant galaxies usually merge after higher evolution (700-750 Myr). We can use this model until both galaxies merge. 
The spatial information should be maintained in shape with optimum zoom, so the zero motion energy cannot go off the frame. We can perform both classification and regression for determining the dynamical parameters. 
Our first approach was to perform the classification of images into several classes. In supervised machine learning, the number of output classes can be infinite or finite. According to \citet{duda2006pattern}, it can be regarded as a classification problem if the output classes are limited. As the GalMer database simulates only a few inclination values, spins also have discrete values, and our classes include discrete intervals in viewing angle and azimuthal angle. This finite number of classes allows us to use DCNN based classification. 

Our second approach was to perform regression to determine dynamical parameters only if we could achieve good accuracy for minute classification in each dynamical parameter. As $\theta$ and $\phi$ are continuous variables, we can perform regression on these parameters effortlessly, but ($i$) and spin value are discrete, requiring classification. In the real-world scenario, we can have the continuous value of inclination. It makes our model more dynamic and efficient to predict such inclination values on which model is not even trained. Spin value is limited to binary class so that the model can predict the same spin of the galaxy.

As \citet{duda2006pattern} mentioned, massive data is necessary to train DCNN to attain a desirable accuracy. It is not always true, but Deep learning requires huge training data because of the significant number of parameters needed to be tuned by a learning algorithm. Deep learning starts with a poor initial state, and then some gradient-based learning algorithm is used to converge the network to an optimal solution. This process requires an enormous amount of data. Also, massive data is not enough; the data should be consistent. DCNN extracts feature only based on image pixels; data should represent a single phenomenon, sequence, or pattern. For our DCNN, we have considered interactions of the mass ratio of 1:1, and between morphological types, we have used gSa and gSb only at their pericentric approach.
We performed several experiments on the dynamical parameters based on the GalMer environment for spin, $i$, $\theta$, and $\phi$. During our experimental process, we considered discrete classes for inclination values of 0°, 45°, 75°, and 90°, and we tried to refine our classes from the limit of each 45\textdegree change for $\phi$ and $\theta$. However, we
realized that our classes could be more refined after increasing the training data set. Hence, the classification model has 60 different classes, which included four discrete values of ($i$), two other classes for a spin (prograde and retrograde), every 5\textdegree change for $\phi$ belonging from 0° to 180°, and $\theta$ belonging from 0° to 90°. This classification can give angles lying in a particular range of degrees. To get the accurate value of each parameter, we proceed further to regression for each dynamical parameter. Here we have to feed four values with corresponding images to the regression model. These four values correspond to the same parameters used in classification. 
There are in total nine dynamic parameters, but we have focussed on four parameters spin, relative inclination ($i$), viewing angle ($\theta$), and azimuthal angle ($\phi$) in our work because of its undue importance in a galaxy merger. We are not planning to classify additional parameters as we used are governing parameters to prepare dynamic models of interacting galaxies. Still, we can deploy our model to those parameters by increasing target classes for regression and classification. \\

\hspace{-12px}
\textbf{Significance of parameters in dynamical models of interacting galaxy pairs}\\

The ($i$), ($\theta$), and ($\phi$) can determine the overall geometry of the interacting pair of galaxies. ($i$) is the angle between the discs of the galaxies. In the GalMer database, ($i$) is the angle of the first galaxy with respect to the second galaxy, while the first galaxy is in the orbital plane Section \ref{Galmer Database:2}. Hence, it is considered a dynamical parameter, so it is responsible for the structure and dynamics of the system. As mentioned in Section \ref{Introduction}, \cite{toomre1972galactic} used simplistic test particle simulation to show the value of inclination. The spin of the galaxy is classified as Prograde and Retrograde. When the galactic spin is aligned with orbital motion, it is called the prograde spin of the interacting pair, whereas retrograde are oppositely aligned.

\citet{di2007star} show that retrograde encounters have greater star formation efficiency than prograde encounters. A retrograde passage exhibits different behavioral characteristics compared to prograde. As mentioned in Section \ref{Galmer Database:2}, the GalMer database firmly supervises the creation of tidal attributes in interacting galaxies. For an equal mass encounter (1:1 mass ratio encounter in GalMer database), a prograde method leads to the generation of curved and long tails for the smaller value in inclination ($i$). The formation of long tidal bridges results in prograde passage of unequal mass encounters (1:2 and 1:10 mass ratio in GalMer database). In the retrograde passage, tidal features may not be developed well. From this, we can know the dynamic significance of the spin and ($i$) in controlling the comprehensive morphology of an interacting pair of the galaxy. Although galactic discs are self-gravitating, the results might diverge in a small amount in the actual scenario. Also, the relative inclination to the orbital plane is not the same as the relative inclination between the galaxies of an interacting pair. The observed image of an interacting pair of the galaxy does not represent the inclination of every galaxy regarding the orbital plane of the sky, which is normal to the observer's optical axis. The proper arrangement with the orbital plane leads to the evolution of well-defined tidal attributes instead of the case in which the galaxy plane is normal to the orbital plane. To put it another way, the details of this galaxy morphology extracted from the observed images can be used as a pointer of the relative angle of inclination with the orbital plane of the galaxy.

As said in Section \ref{Galmer Database:2}, in the plane of galaxy interacting pairs, the viewing angle is the first galaxy plane. The observational parameter that controls the geometry of the galaxy pair interaction is projected in the sky and has no impact on the system's dynamics. Like, the tidal features, when projected on the sky plane, its length may appear different for the observer. From this, we can say that observed images of interacting pairs of galaxies have an essential role of $\theta$ for determining an accurate dynamical model.

\section{Data Extraction and preprocessing}
\label{Data collection and preprocessing:4}

In our present work, as mentioned in section \ref{Galmer Database:2}, we used a 1:1 mass ratio galaxy interaction (giant Sa type spiral galaxy(gSa) and giant Sb type spiral galaxy(gSb)). Here we need to collect images manually, but we developed a web scraping method employed in \citet{singh2020classification} for bulk downloading images of 400×400 in GIF (Graphics Interchange Format). After downloading the original images, we rescaled the image size to 64×64 and converted it in JPG (Joint Photographic Experts Group) format. We executed our model with the original 400×400 images and a moderate resolution of 128×128 and 256×256. The model accuracy slightly decreases with lowering the resolution but, an increase in the dataset was more feasible than high-resolution images, so we considered low-resolution images. Low-resolution even helps store more images in the available ram allowing us to train a massive number of data with larger batch size. 

Even after bulk downloading the images, it is still insufficient to train DCNN. To overcome the shortage of data, we performed augmentation on the extracted images. We have used conventional image augmenting operations on the primary images, resulting in the dataset's increased size. It includes rotation, intensity variation, and gaussian blur \citet{krizhevsky2012imagenet}, \citet{almasi2016review}. In the database formed after augmenting, we have split 80\% of the total number of images for training, and the remaining 20\% of the dataset, we used 10\% for validation and 10\% for testing.

\section{Deep Convolutional Neural Networks}
\label{Convolutional Neural Networks:5}

The basic structure of artificial neural networks (ANNs) consists of connected artificial neurons. Each neuron is characterized by an activation function, which acts on the input. Neural networks have 3 layers: input, hidden, and output layers. The role of ConvNET is to transform the images in a simple form, which is feasible to process without any loss in the main features that are important for a good prediction. The kernel window shifts from left to right over the image and performs a matrix multiplication operation in consecutive network layers. The central concept of CNN is to adopt an architecture that extracts high-level features from an input image. As CNN gives high accuracy and is less complicated, we have used it in the classification and regression of our dynamical parameters. Hyperparameters like learning rate, number of epochs, batch size, activation function, number of hidden layers, dropout are needed to configure for optimum CNN model. We have used one of the well-known architectures, ALexNet, because of its simplicity to detect millions of objects, minimizing overfitting. It was also used in \citet{krizhevsky2012imagenet}, which gave them exceptional results.

We have modified the AlexNet design architecture mentioned in table\ref{CNN}, consisting of 12 layers of convolution 2-D, 5 layers of max-pooling, and dropouts. Each layer has a pooling window size of (2,2) and padding of the same size. In the end, it is connected to fully connected layers. Max pooling layers reduce the dimensions and the computation power required, making the model more robust and precise. We used dropout layers with a probability of variations 0.3 to avoid overfitting \citet{srivastava2014dropout}. The activation function transforms all the weighted inputs to the outputs. So for all the layers, we preferred ReLu (\citet{nair2010rectified}), which returns the positive input directly and converts the negative values to zero. This makes it easy to train our model and achieve better results. The last layers are fully connected, which gather data from the previous layers and form the final output, followed by a dense final layer with a sigmoid activation function, which introduces non-linearity in our neural network model and gives output between 0 to 1 (\citet{han1995influence}.

\begin{table}
\caption{Architecture for CNN classification}
\scalebox{0.9}{
\begin{tabular}{|c|c|c|c|}
\hline 
\label{CNN}


\textbf{Layer Number} & \textbf{Type of Layer} & \textbf{No. of Filters} & \textbf{Parameters} \\
\hline
1 & Convolutional & 16 & 208 \\
\hline 
2 & MaxPooling & 16 & 0 \\
\hline
3 & Convolutional & 64 & 4160 \\
\hline 
4 & MaxPooling & 64 & 0 \\
\hline 
5 & Convolutional & 128 & 32896 \\
\hline 
6 & MaxPooling & 128 & 0 \\
\hline 
7 & Convolutional & 256 & 131328 \\
\hline 
8 & MaxPooling & 256 & 0 \\ 
\hline 
9 & Convolutional & 512 & 524800 \\
\hline 
10 & Convolutional & 512 & 1049088 \\
\hline 
11-13 & ... & ... & ... \\
\hline 
14 & Convolutional & 512 & 1049088 \\ 
\hline 
15 & Convolutional & 256 & 524544 \\
\hline 
16 & MaxPooling & 256 & 0 \\
\hline 
17 & Convolutional & 128 & 131200 \\ 
\hline 
18 & Dense & 128 & 16512 \\ 
\hline
19 & Fully Connected & 512 & 0 \\
\hline
20 & Dense & 60 & 30780 \\
\hline
\end{tabular}}
\end{table}

\subsection{CNNs in Image regression}
\label{CNNs in Image Regression}
\subsubsection{AutoKeras}

AutoKeras \citet{jin2019auto} is a free AutoML system based on Keras. DATA Lab developed it at Texas A \& M University. The idea behind AutoML is to minimize human participation in building models instead of developing one's model architecture and tuning the parameters aiming at the best results. At the same time, the manual approach to machine learning assumes the whole model development pipeline to be made by a human-machine learning expert. Using the AutoKeras library, we are firmly able to access deep neural networks. Keras library and Python programming language are used to develop this software. Using AutoKeras locally on its machine instead of configuring Docker and Kubernetes in the cloud is its main benefit. More expansive search space to the recurrent neural network is planned in a future release to solve computer vision tasks.

It aims to give us promising results by choosing the best neural architecture, finding an effective learning algorithm, and optimizing the parameters for the assigned dataset. AutoKeras is an open-source library. AutoKeras offers a neural architecture search algorithm that contains Bayesian Optimizer and Gaussian Process, a module defined as a Searcher. These algorithms utilize the CPU. As we all know, the system utilizes GPU for training the model; for this, the model
trainer module is defined. Here, it trains the neural network with the training data in a separate process for both GPU and CPU to get utilized simultaneously. The searcher for processing computational graphs controls the Graph module. The current neural architecture in the graph uses RAM for faster access, which gives efficient results and reduces time complexity. As we know, the size of the Neural Network is significant, and is not easy to store all of it on RAM. The model saves the trained models on the storage devices.
Moreover, the AutoKeras library's advantage is restoring the previous weights. As we train deep neural networks, there might be a possibility for the process to stop, and it kills the time to train the model again. Nevertheless, AutoKeras saves the model and the trained parameters, weights of every epoch, and even the best-trained model on the storage device.

\subsubsection{Residual Network Design}
\label{Resnet-network}
The ResNet architecture has revolutionized the deep learning neural network. The deeper the network better the accuracy is expected. However, as mentioned in the paper \citet{he2016deep}, it is observed that if we increase the depth of architecture steadily, it reaches its optimum training error and then degrades the accuracy. Deep Residual Network has solved this vanishing gradient problem.
ResNet first introduced the skip connection concept. In this method, we add the original input to the output of the block while we stack the convolution layers to increase the complexity of the network. Skip connection is applied before the Relu activation to obtain the best results. The skip connection allows an alternate shortcut path for information to flow from earlier layers into the model of later layers, solving the vanishing gradient problem. If the input and output dimensions are the same, we can add these identity shortcuts to the network. Nevertheless, if the dimensions are not similar, padding can be done with extra zeros to increase the dimensionality of the layer, and the shortcut connection can perform the identity mapping. Using ResNet 50, we can improve the depth of architecture, but simultaneously, we can achieve an accuracy as there is a 20.74\% decrease in training error compared to plain neural networks.
It consists of convolution and identity blocks, and each block contains three convolutional layers. The ResNet-50 model consists of 5 stages, each with a convolution and Identity block. Each convolution block has three convolution layers, and each identity block also has three convolution layers. Hence ResNet gives exceptionally better results.

\begin{figure}
\centering
\includegraphics[width=250pt,height=600pt]{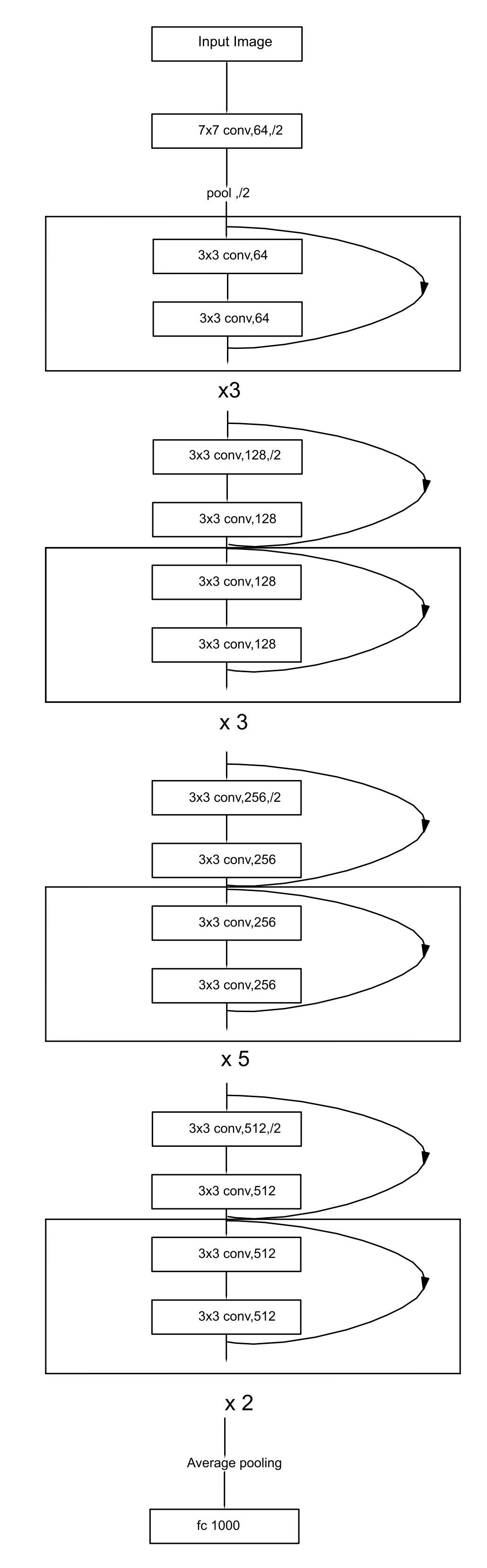}
\caption{Residual network with 50 layers and more than 23 million parameters used for regression.}
\label{resnet-arc}
\end{figure}

\section{Results, Validation and Discussion }
\label{Results}

By using Keras library(\citet{chollet2015keras}), we have implemented our model in Python. Running a ConvNet Deep Learning model requires heavy usage of GPU and RAM, we used Google Colab pro as it was the best source available for us to get results efficiently. The specifications of the server which we used are: \newline
GPU: 1xTesla V100-SXM2, compute capability 6.0 , having 5,120 CUDA cores, 16 GB GDDR5 VRAM, 1.53 GHz Frequency ,300 Wattage  \newline
CPU: 2x single-core hyperthreaded Xeon Processors @2.3 GHz, i.e. (2 core, four threads)\newline
RAM: 25.46 GB Available \newline

\subsection{Classification of dynamical parameters}

We performed a classification of 60 classes on the spin, $i$, $\theta$, and $\phi$. The classification includes two classes of prograde and retrograde spin, four classes of the inclination of angles of 0°, 45°, 75°, and 90°. The remaining classes are the combinations of angles between viewing and azimuthal angles mentioned in \ref{CNNs in dynamical parameter estimation:3}. The images are scaled down to 64×64, as discussed in Section \ref{Data collection and preprocessing:4}. It takes around 3 minutes, 7 seconds to run the first epoch and less than 3 minutes for consecutive epochs with a learning rate of 0.001 using NVIDIA CUDA Deep Neural Network library (cuDNN). For 15 epochs, the powerful GPU finishes computing in just 45 minutes. \\

\begin{figure}
\centering
\includegraphics[scale=0.6]{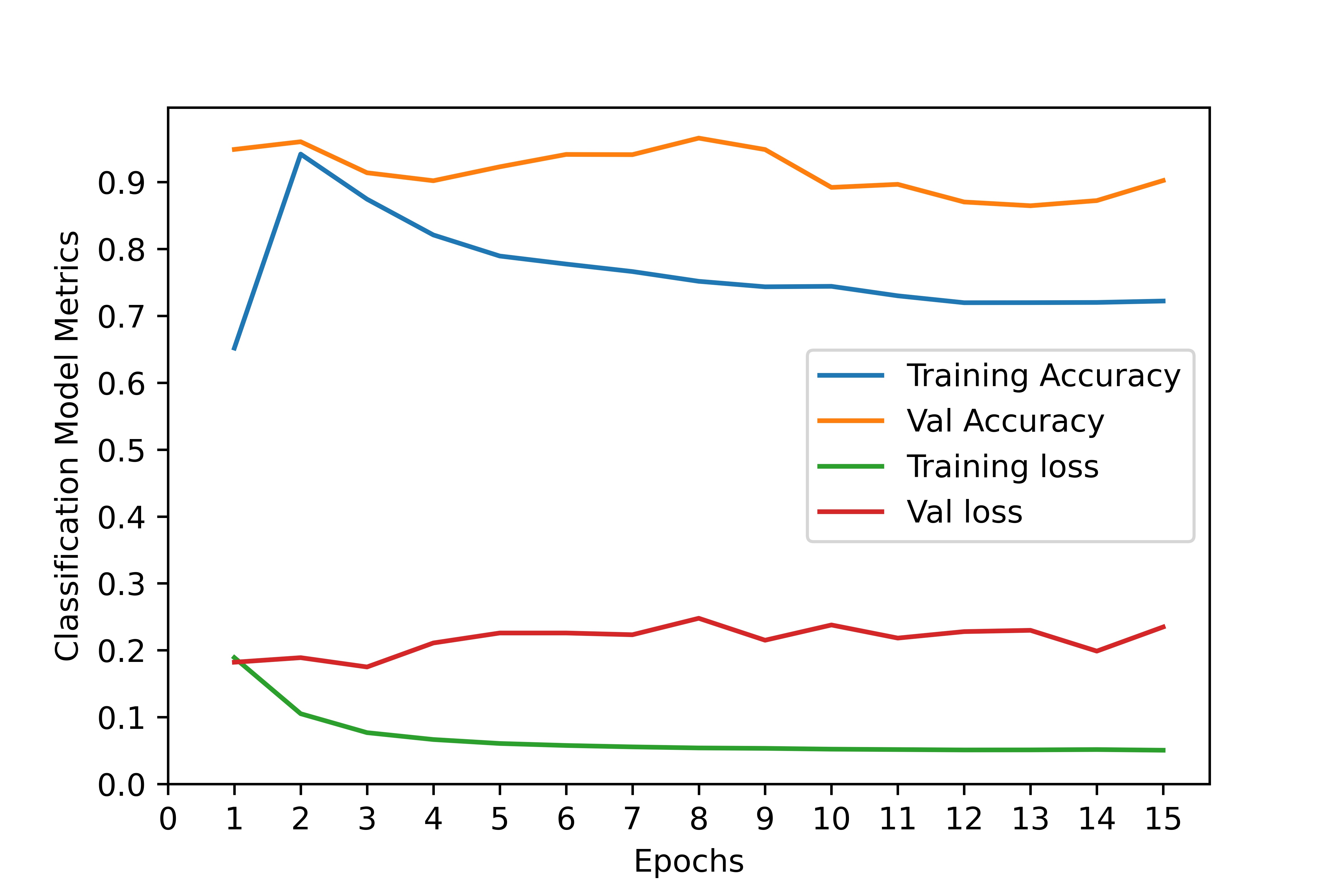}
\caption{Performance of Classification Model. The Y-axis shows the classification model metrics, while X-axis shows the number of iterations completed during training (Epochs). The model can achieve 93.63\% in just the second epoch.}
\label{fig:Learning curve for classification}
\end{figure}

As mentioned above, learning curves are illustrated in fig \ref{fig:Learning curve for classification}, which represents the training progress of different cases. In fig \ref{fig:Learning curve for classification}, we have shown the behavior and progress of accuracy, loss function as the number of epochs increases. It can be noticed that we achieved an accuracy of 93.63\% just after the second epoch. Here we can see that the validation accuracy is more than the training accuracy. A Keras model has two modes: training and testing. Regularization mechanisms, such as Dropout and L1/L2 weight regularization, are turned off at testing. They are reflected in the training time loss but not in the test. We verified our data on the same model without adding any dropout layers, which resulted in a slight decrease in accuracy, but testing accuracy was less than the model accuracy. Besides, the training loss that Keras displays are the average of the losses for each batch of training data over the current epoch. Because the model changes over time, the loss over the first batches of an epoch is generally higher than over the last batches. This can bring the epoch-wise average down. On the other hand, the testing loss for an epoch is computed using the model at the end of the epoch, resulting in a lower loss.

\subsection{Regression of dynamical parameters}
\begin{figure}
\centering
\includegraphics[width = 0.4\textwidth]{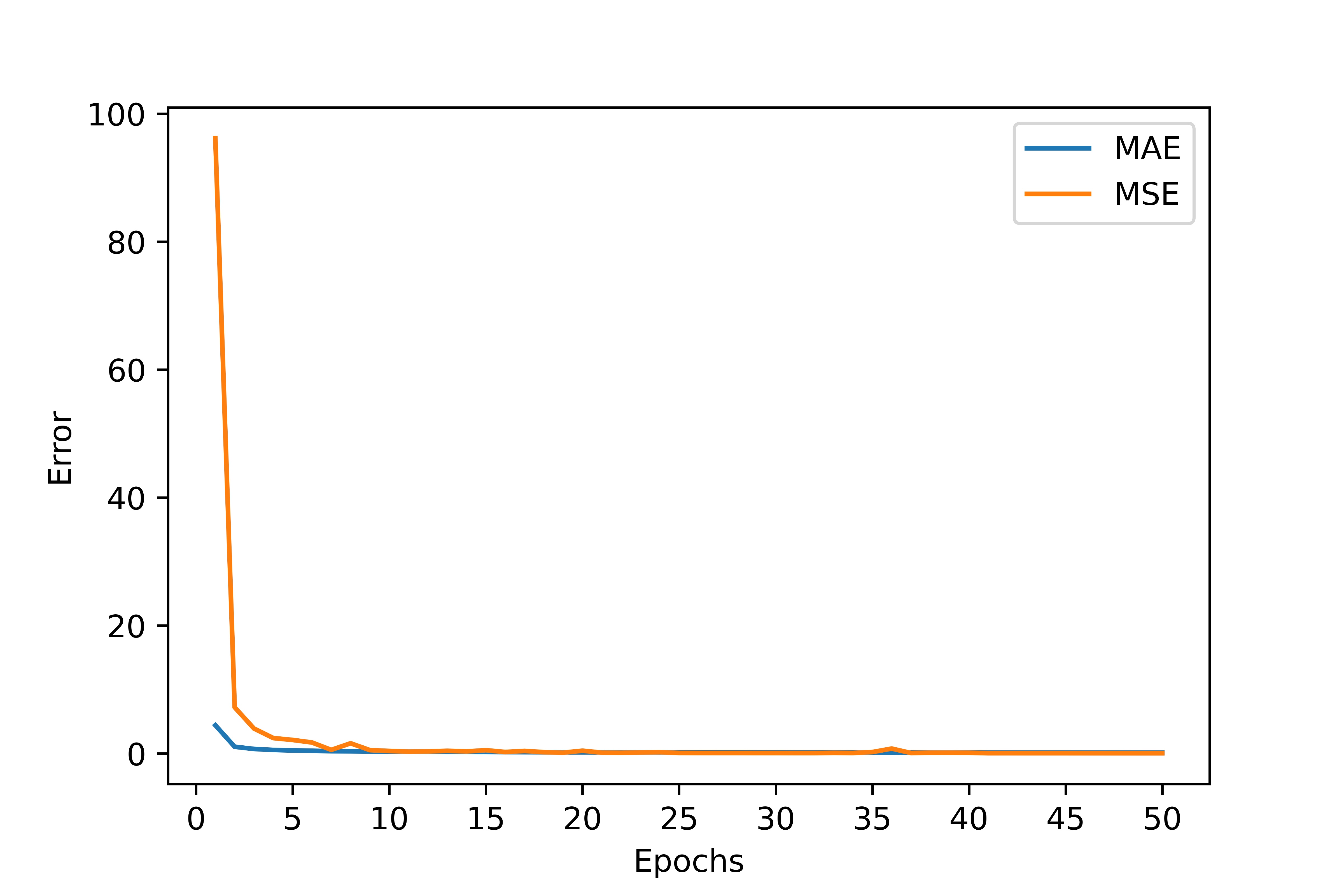}
\caption{Performance of Regression Model with an increase in the number of epochs during training. The Y-axis shows the mean absolute and squared errors, while the X-axis shows the number of iterations completed in training (Epochs). We observe that as the number of epochs increases, the mean absolute error and mean squared error decreases.}
\label{fig:Epochvserror}
\end{figure}

As discussed in Section \ref{CNNs in Image Regression}, we have the discrete value of $i$ and continuous values of $\Phi$ and $\theta$, so effortlessly, we can use CNN-based regression architecture. To estimate the accurate value of the parameters of interacting galaxies, we have used ResNet50 (Section \ref{Resnet-network}) architecture consisting of 50 layers. We have used the Adam optimization algorithm (\citet{kingma2014adam}) to minimize the loss function, mean absolute error as loss, and metrics of mean squared error, with a learning rate of 0.001. We have trained the model for 50 epochs. and the results are shown in the  Figure \ref{fig:Epochvserror}. We have used 2,63,536 sample data ranging from all spin, inclination, theta, and phi values. For our experimentation, we have chosen image samples corresponding to gSa and gSb interactions (at their pericentric approach). However, experiments can be executed on other types of interactions. We have used 2,37,182 samples for training and 26,354 samples for validation. After each epoch, we computed loss (mean absolute error) and mean squared error (MSE) on both training and validation data. We found that the loss and MSE values are initially high, but we can observe a slow and gradual reduction as the training progresses. Due to the high computational cost for each epoch, we have stopped training after 50 epochs, as it has not reached the overfitting limit. Loss and MSE values could be decreased in the future. If we train the model for more epochs, it can perform better. \\

Evaluation of the model performance has been done using the test data. The performance is expressed using R-squared, Mean Square Error(MSE). The goodness of fit of a regression model is measured using R-squared, which is a statistical measure. It measures the proportion of the variability of the linear relationship between the dependent and independent variables. It compares the residual sum of squares ($SS_{res}$) with the total sum of squares($SS_{tot}$). The model is fitted better if the R-squared value is closer to 1. The mean squared error (MSE) tells you how close the predicted values are to the actual value by measuring the distances from the points to the regression line and squaring them. It tells the model flaws as it gives more weight to larger differences.

\begin{align}
\hspace{46px}   
R^2 = 1 - \frac{SS_{res}}{SS_{tot}} 
\end{align}

\begin{align*}
 where, \hspace{5px} SS_{res} &= (y - \hat{y})^2 \\
 \hspace{5px} SS_{tot} &= (y - \Bar{y})^2 \\ 
 \hat{y}&= predicted \hspace{3px} value  \\
  \Bar{y} &= mean \hspace{3px} of\hspace{3px}all\hspace{3px}y\hspace{3px}values  \\
  y &= actual \hspace{3px} value
\end{align*}

\vspace{-10px}

\begin{align}
\hspace{34px}   
MSE =\frac{1}{n} \sum\limits_{i=1}^{n} (Y_i - \hat{Y}_i)^2
\end{align}

\vspace{-8px}

\begin{align*}
\hspace{49px} 
        n  &= number \hspace{3px} of \hspace{3px} data \hspace{3px} points , \\
      Y_i  &= observed \hspace{5px} value, \\
\hat{Y}_i  &= predicted \hspace{5px} value.
\end{align*}

\begin{figure}
\centering
\includegraphics[width = 0.4\textwidth]{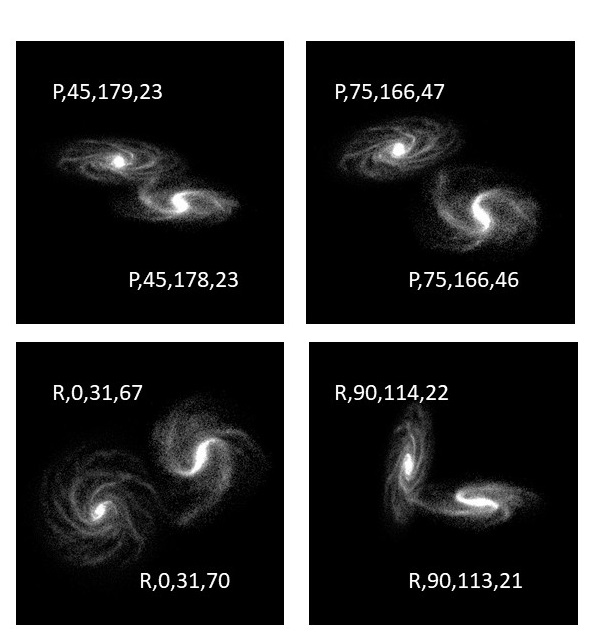}
\caption{Sample prediction of the regression model. The top left corner illustrates the actual values of dynamical parameters, and the bottom right corner depicts our regression model's predicted values of the interacting galaxies. The parameter spin types are represented as prograde defined by P and retrograde by R. So the series of the parameter is shown as P or R, $i$, $\phi$, $\theta$}
\label{fig:test_result}
\end{figure}

As shown by (\citet{duda2006pattern}) to achieve good generalization capability, the model should display low error on testing data, although the training error can be higher. Finally, we obtained an MAE of 0.1301 and MSE of 0.0503 on training data and obtained an MAE of 0.2012 and MSE of 0.0833 ($\approx 0$) on testing data. That means the network has achieved a good generalization performance. Here we have achieved an $R^2$ value of 0.9986. Therefore the model is unbiased and has a minor variance. Fig \ref{fig:test_result} represents the results of a few samples from our testing data that is fed into our regression model.

We note here that spin is the classification of prograde and retrograde, but our model is highly efficient to predict accurate values of prograde and retrograde as corresponding 0 and 1. As data available in GalMer database, $i$ is a discrete set of 0\degree, 45\degree, 75\degree, 90\degree, our model can predict the nearest value to the inclination. In the real-world scenario, we can have any inclination value, and our model efficiently predicts that inclination. The remaining $\theta$ and $\phi$ values are continuous variables so that the prediction will be close to the actual $\theta$ and $\phi$. Therefore, we have developed a fully-fledged regression model and it can determine all the four dynamical parameters for any given image of an interacting galaxy pair.\\

\hspace{-11px}
\textbf{Results on SDSS data} \\

\hspace{-11px}
As we have simulated our DCNN model using N-body + Smoothed-particle hydrodynamics galaxy merger simulated data. Our trained DCNN model is verified on actual data available on Sloan Digital Sky Survey (SDSS) Data Release 16 (DR16). Data Release 16 (DR16) is the fourth data release of the fourth phase of the Sloan Digital Sky Survey (SDSS-IV). DR16 contains SDSS observations through August 2018 and from all data from the earlier releases. We can collect image information using SQL queries depending on declination (DEC) and right ascension (RA) parameters on SDSS CasJobs. The SDSS has actual galaxy merger data; it is different from the simulated data and has distinct features.

For instance, images obtained from SDSS do not provide us any information about the galaxies orbital planes. However, intrinsically, viewing angle and inclination is the same for SDSS images. There are chances that they might exhibit distinct characteristics or phenomena from images of the GalMer database. We have tested our model on a few images to verify the prediction capability of the network. We have used the same testing data used in (\citet{prakash2020determination}). To get the data set, we have labeled the images manually. The chosen data from SDSS has been examined visually for nature and testing viability. To extract the independent value of position angle and inclination for every galaxy from the interacting pair, we used galaxies and a cosmology database named HyperLEDA. Here, the angle delimited by the galaxy's long axis regarding celestial north is known as position angle, and it is resolved in terms of 0° to 180° from North to East. In HyperLEDA, individual inclinations and position angles of the interacting galaxy pair are indicated by $i_1$ and $i_2$ and $PA_1$ and $PA_2$, respectively. We have used equation \ref{equation} to determine the relative inclination. The data available for testing our model on real images from SDSS contains only relative inclination as a parameter. However, our model is a fully-fledged model which can predict all four parameters mentioned throughout the paper. These testing images are resized to (64,64), and decreasing pixel density made the images unclear. Because of these image transformation activities, errors for predictions increased significantly. The $R^2$ value for testing data is 80.48\%. Mean absolute error and mean squared error are 5.44 and 46.69, respectively. The value might have improved if we had information about other parameters. The data we have used and the predictions on that data can be accessed in section \ref{Data Availability}.

\begin{multline}
cos(i) = (sin(i_1) sin(i_2) cos(PA_1) cos(PA_2)+ \\
(sin(i_1) sin(i_2) sin(PA_1) sin(PA_2) + cos(i_1) cos(i_2)
\label{equation}
\end{multline}

It is to be mentioned that interacting galaxy images of mass ratio 1:1, taken from the GalMer Database, have been used to train our CNN model. Although, in reality, the mass ratio can achieve any domain of values. So, there might be any possibilities that our model could lag or could give wrong predictions. However, we have just tested on the 1:1, and we could train the model for different mass ratios available on GalMer \ref{Galmer Database:2} and could predict the required predictions. The stellar mass ratios of these interacting galaxies from actual data can be computed using the definite value of each galaxy for every interacting pair of a galaxy in distinct bands spanned a range of values (Appendix \ref{Appendix}). As we tested on 1:1 mass ratios, our model is reasonably successful. We could state that it might also work for different mass ratios, but it is advised to train them for better results.

\section{Conclusion}
\label{Conclusion}

We have illustrated the implementation of Convolutional Neural Networks in astronomy for finding the exact dynamical parameters of the interacting pairs of galaxies. The relative inclination is the angle between the disc of the interacting galaxy. The angle subtended by perpendicular to the orbital plane and line of sight is the viewing angle. Spin is the dynamical parameter that shows in which manner galaxies are merging. Along with other parameters, the azimuthal angle will help determine the galaxy's geometry. We have collected data from the GalMer database for training, which is an N-body + SPH simulation.

We have used images of mass ratio 1:1 of galaxy pair interaction at their pericentric approach. As GalMer provides both discrete and continuous sets of values, we can apply both approaches for determining the dynamical parameter. The training sample represents galaxy interactions between gSa and gSb morphological galaxies, giving many parameters characterizing the dynamical models. Our model is trained on discrete inclination values of inclination values, but it is very robust that it can even detect the galaxy interactions with the inclination values between them (i.e. angle between i = 0; 45; 75; 90). Our model can classify the type of galaxy interactions as prograde and retrograde, making the model very dynamic in terms of classification. The model is trained on the continuous set of theta and phi values. The regression gives promising results with an MSE of 0.0503 and achieved an R squared value of 0.9986. Therefore the regression model is highly unbiased and has a minimum variance. Apart from the data from the GalMer database, we have tested our model on actual data from SDSS DR 16, which can be very useful for determining the dynamical parameters of the galaxy.

\section{Acknowledgements}

We want to thank Mr. Prem Prakash for helping us at many stages where we were facing problems. We would also express our gratitude to the anonymous referee for their remarks and opinions, which helped us gain deep insight into the topic and enhance the standard of our paper.

\section{Data Availability}
\label{Data Availability}
The GalMer simulations can be downloaded at \href{http://www.projet-horizon.fr.}{Project Horizon}. For downloading bulk images, someone can refer to the first step mentioned in \href{https://github.com/JanvitaReddy/Parameters-determination}{Github-Repository}.
For predicting the dynamical parameters using our pretrained model, we can run the code from the second step to the last step mentioned in \href{https://github.com/JanvitaReddy/Parameters-determination}{Github-Repository}. Someone can view the test samples we used for testing our regression model \href{https://github.com/JanvitaReddy/Parameters-determination/raw/master/REAL_DATA_PRED.xlsx}{here}.



\bibliographystyle{mnras}
\bibliography{example} 




\ Appendix

\section{Appendix}
\label{Appendix}

\begin{figure*}
\includegraphics[width = 0.8\textwidth]{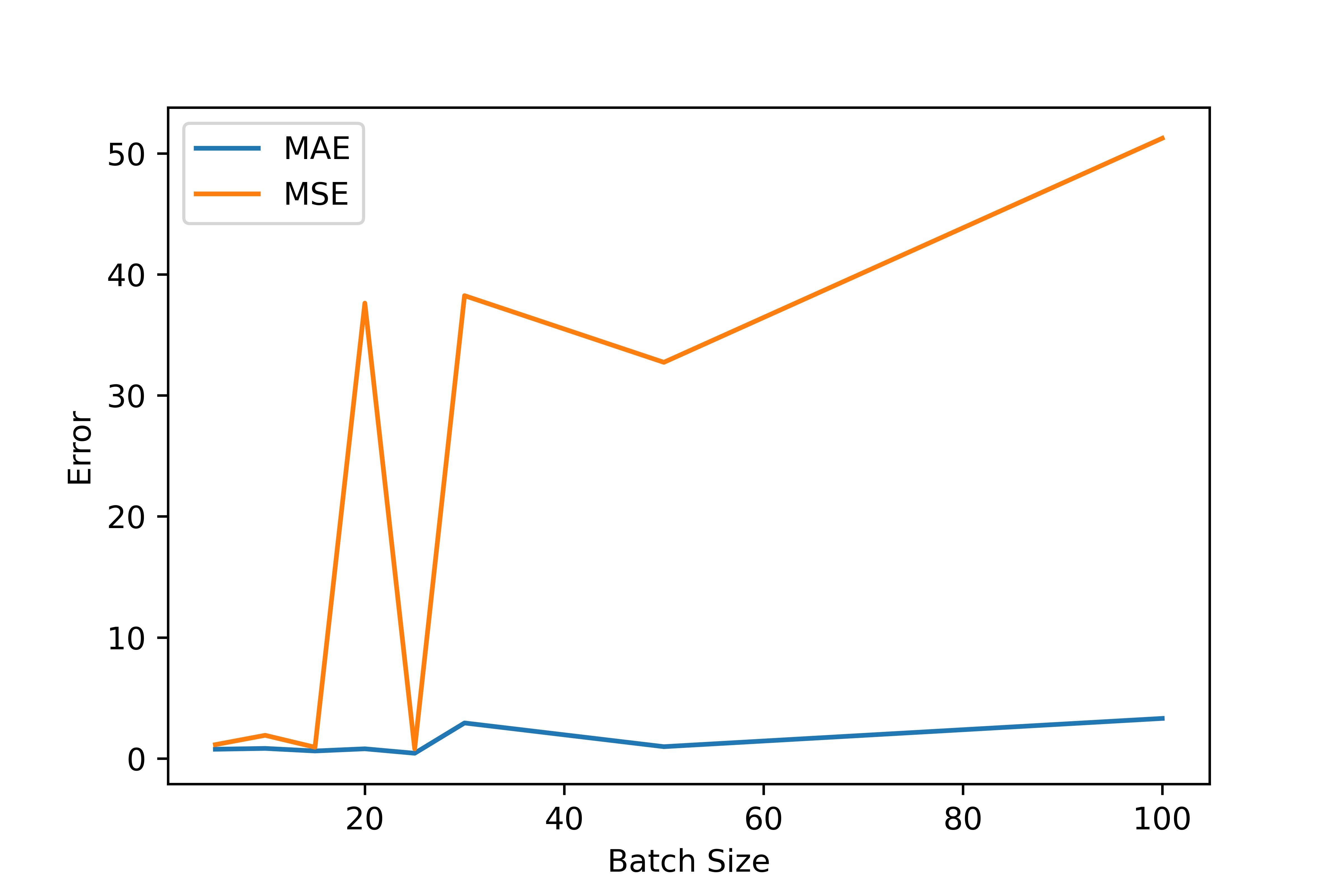}
\caption{Performance of Model with change in the Batch Size during training. The Y-axis shows the mean absolute and squared errors, while the X-axis shows the number of images given at the training time (Batch Size). As we know, lower the batch size lesser the error, but due to the vast data needed to train the model, it was not feasible to take a very low value. The optimum batch size for the model we considered here is 25.}
\label{fig:Batchvserror}
\end{figure*}

\begin{figure*}
\includegraphics[width = 0.8\textwidth]{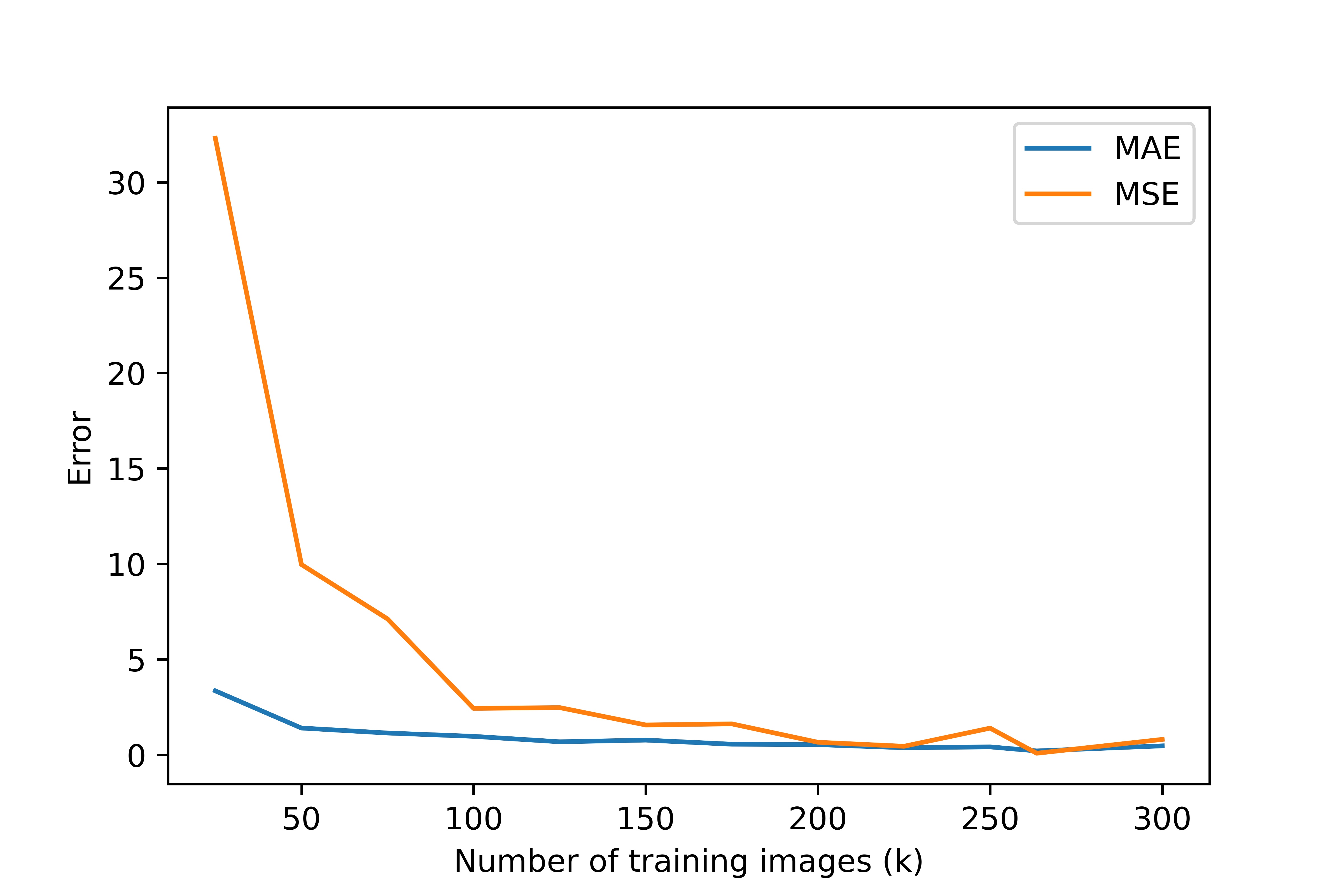}
\caption{Performance of the model with an increase in the training data. The Y-axis shows the mean absolute and squared errors, while the X-axis shows the Number of Images used to train the model. We observe that as we increase the training dataset, the error decreases.}
\label{fig:trainingvserror}
\end{figure*}

\begin{figure*}
\centering
\includegraphics[width = 0.9\textwidth]{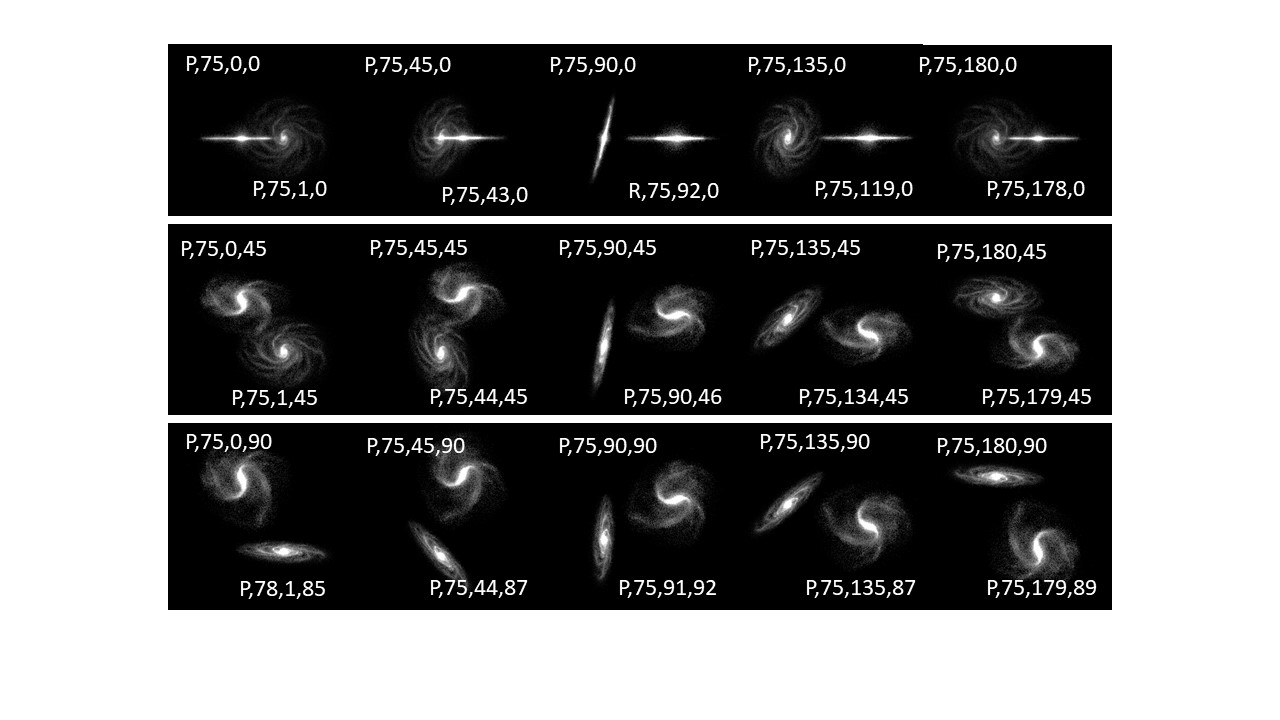}
\caption{ Montage of sample images for regression. It represents interacting galaxies, having prograde spin, an inclination of 75°, ($\phi$) ranging from 0° to 180°, and ($\theta$) ranging from 0° to 90°. Each galaxy interaction image's top left and bottom right represents the actual and predicted dynamical
parameters.}
\label{fig:FigureA6}
\end{figure*}

\begin{figure*}
\centering
\includegraphics[width = 0.9\textwidth]{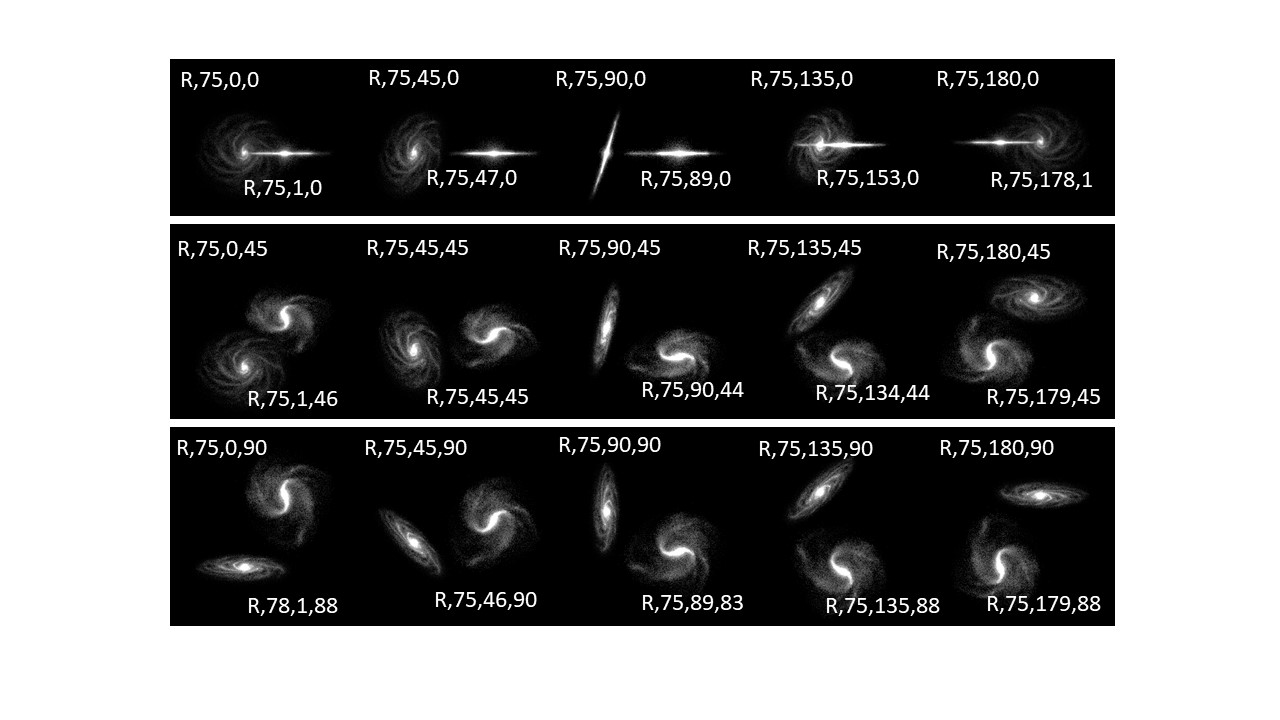}
\caption{ Montage of sample images for regression. It represents interacting galaxies, having retrograde spin, an inclination of 75°,($\phi$) ranging from 0° to 180°, and ($\theta$) ranging from 0° to 90°. Each galaxy interaction image's top left and bottom right represents the actual and predicted dynamical
parameters.}
\label{fig:FigureA10}
\end{figure*}

Montages of some images have been shown in  Figures \ref{fig:FigureA6},\ref{fig:FigureA10}. In these figures, ($i$) has a discrete value of 75\textdegree, ($\theta$) ranges from 0\textdegree to 90\textdegree represented along Y-axis, and ($\phi$) ranges from 0\textdegree to 180\textdegree represented along the X-axis. The top left corner illustrates the actual values of dynamical parameters, and the bottom right corner depicts our regression model's predicted values of the interacting galaxies. The parameter spin types are represented as prograde defined by P and retrograde by R. So the series of the parameter is shown as P or R, $i$, $\phi$, $\theta$

\bsp	
\label{lastpage}
\end{document}